# Large-composition-range pure-phase homogeneous InAs$_{1-x}$Sb$_x$ nanowires

Lianjun Wen,[†,‡] Dong Pan,[*,†,‡,⊥] Lei Liu,[†,‡] Shucheng Tong,[†] Ran Zhuo,[†,‡] and Jianhua Zhao[*,†,‡,§]

[†]*State Key Laboratory of Superlattices and Microstructures, Institute of Semiconductors, Chinese Academy of Sciences, P.O. Box 912, Beijing 100083, China*

[‡]*College of Materials Science and Opto-Electronic Technology, University of Chinese Academy of Sciences, Beijing 100049, China*

[⊥]*Beijing Academy of Quantum Information Sciences, 100193 Beijing, China*

[§]*CAS Center for Excellence in Topological Quantum Computation, University of Chinese Academy of Sciences, Beijing 100049, China*

(Dated: November 3, 2021)

**Narrow bandgap InAs$_{1-x}$Sb$_x$ nanowires show broad prospects for applications in wide spectrum infrared detectors, high-performance transistors and quantum computation. Realizing such applications require the fine control of composition and crystal structure of nanowires. However, to date, the fabrication of large-composition-range pure-phase homogeneous InAs$_{1-x}$Sb$_x$ nanowires remains a huge challenge. Here, we first report the growth of large-composition-range stemless InAs$_{1-x}$Sb$_x$ nanowires ($0 \leq x \leq 0.63$) on Si (111) substrates by molecular-beam epitaxy. It is found that pure-phase InAs$_{1-x}$Sb$_x$ nanowires can be successfully obtained by controlling the antimony content *x*, nanowire diameter and nanowire growth direction. Detailed EDS data show that the antimony is uniformly distributed along the axial and radial directions of InAs$_{1-x}$Sb$_x$ nanowires and no spontaneous core-shell nanostructures form in the nanowires. Based on field-effect measurements, we confirm that InAs$_{1-x}$Sb$_x$ nanowires exhibit good conductivity and their mobilities can be up to 4200 cm$^2$V$^{-1}$s$^{-1}$ at 7 K. Our work lays the foundation for the development of InAs$_{1-x}$Sb$_x$ nanowire optoelectronic, electronic and quantum devices.**

*E-mails: pandong@semi.ac.cn; jhzhao@semi.ac.cn

Ternary III-V nanowires, whose properties can be tuned and designed through varying composition, have aroused tremendous research interest in recent years.[1-9] Among them, InAs$_{1-x}$Sb$_x$ nanowires are of particular interest due to their tunable narrow bandgaps (0.1 ~ 0.4 eV), high electron mobilities, strong spin-orbit interactions and large g-factors.[10-15] These unique properties make them an ideal platform for studying wide spectrum infrared detectors, high-speed transistors and topological physics.[2, 7-11, 16-20] To realize these applications, much work has been done to grow InAs$_{1-x}$Sb$_x$ nanowires on various substrates by chemical vapor deposition, metal-organic vapor phase epitaxy, chemical beam epitaxy and molecular-beam epitaxy (MBE).[9, 14, 21-32] Nevertheless, there are still many unsolved problems related to the controllable growth of InAs$_{1-x}$Sb$_x$ nanowires.[21, 24-25, 33] On one hand, to date, it is still difficult to fabricate stemless InAs$_{1-x}$Sb$_x$ nanowires with higher antimony contents ($x > 0.35$).[25] On the other hand, these nanowires typically exhibit random mixtures of zinc-blende (ZB) and wurtzite (WZ) phases.[21, 23-26] High-density planar defects, such as stacking fault (SF), rotational twin and grain boundary (GB), are usually found in these nanowires. Such defects can degrade their optical and electrical properties.[34-35] Furthermore, the unevenly compositional distribution has also been found in InAs$_{1-x}$Sb$_x$ nanowires, and even the spontaneous core-shell structure is always observed.[9, 20, 25, 33] These unexpected results are hindering the research of bandgap engineering and topological physics based on InAs$_{1-x}$Sb$_x$ nanowires.[35-36] To the best of our knowledge, up to now, the fabrication of large-composition-range pure-phase stemless InAs$_{1-x}$Sb$_x$ nanowires with homogeneous compositional distribution has not been realized yet.

In this work, we demonstrate the growth of large-composition-range pure-phase InAs$_{1-x}$Sb$_x$ nanowires ($0 \leq x \leq 0.63$) on Si (111) substrates using silver as catalysts by MBE. Detail structural studies confirm that pure WZ or ZB phase InAs$_{1-x}$Sb$_x$ nanowires can be obtained by controlling the antimony content, nanowire diameter and nanowire growth direction. Energy dispersive spectrum (EDS) analyses verify that the antimony is uniformly distributed throughout the nanowire. Electrical measurements show that the grown InAs$_{1-x}$Sb$_x$ nanowires exhibit a high electron mobility.

**Large-range antimony content tunable.** As mentioned above, so far, it is still difficult to prepare large-composition-range stemless InAs$_{1-x}$Sb$_x$ nanowires no matter what the growth manners (foreign-catalyst-free or gold-assisted) were used.[9, 21, 24-25, 29, 33] To resolve this problem, here we choose silver as catalysts to grow InAs$_{1-x}$Sb$_x$ nanowires. We find that large-composition-range InAs$_{1-x}$Sb$_x$ nanowires ($0 \leq x \leq 0.63$) can be successfully grown directly on Si (111) substrates by tuning the antimony flux ($P_{Sb}$). As shown in Figure 1a, with increasing $P_{Sb}$, more antimony atoms can be incorporated into nanowires, thereby obtaining nanowires with higher antimony contents (the corresponding scanning electron microscope (SEM) images can be found in Figure S1 in the supporting information). The antimony content $x$ can be tuned from 0 to 0.63, which is approximately twice comparing with the previously reported results ($0 \leq x \leq 0.35$).[9, 25] We also note that the antimony content $x$ gradually becomes saturated with the increase of the $P_{Sb}$ (see Figure 1a). Meanwhile, the density of nanowires begins to decrease (see Figure S1). Therefore, it is difficult to obtain InAs$_{1-x}$Sb$_x$ nanowires with the antimony content $x$ higher than 0.7.

It is well known that the surfactant effect of antimony plays an important role in the growth of III-Sb nanowires.[4, 14, 21, 24-25, 37-38] It often enhances the lateral growth and inhibits the axial growth of nanowires, which makes nanowires typically show a low aspect ratio (AR) (~ 10). However, we find that our InAs$_{1-x}$Sb$_x$ nanowires still exhibit an ultrahigh AR (> 30) although the surfactant effect of antimony cannot be avoided during nanowire growth. Figure 1b shows a curve of the average diameter and length of InAs$_{1-x}$Sb$_x$ nanowires as a function of the antimony content $x$ (the average diameter is defined as the average of the diameter of the bottom of the nanowire and the diameter of the top of the nanowire). We can see that all InAs$_{1-x}$Sb$_x$ nanowires ($0 \leq x \leq 0.63$) show small diameter (10 ~ 80 nm) and ultrahigh AR (> 30), which is beneficial for quantum device research.[39] It can also be seen that as the antimony content $x$ increases, due to the surfactant effect of antimony, the average diameter and length of nanowires begin to increase and decrease, respectively. The nanowires with a lower antimony content ($x \leq 0.3$) exhibit an ultrahigh AR of over 50, which is 2-4 times higher than the results of foreign-catalyst-free InAs$_{1-x}$Sb$_x$ nanowires.[21, 25, 27, 33]

**Crystal structure and phase purity control.** III-V nanowires typically exhibit a mixture of WZ and ZB phases, and a large number of planar defects usually formed in nanowires.[21, 24-26] Similar results are also observed in our InAs$_{1-x}$Sb$_x$ nanowires with low antimony content ($x < 0.2$). As shown in Figure 2a,b, GBs and SFs can be easily found along the axial direction of nanowires. Such defects will degrade the properties of nanowire devices, so the realization of pure-phase InAs$_{1-x}$Sb$_x$ nanowires is of great significance.[35] Based on detailed structural studies, we find that pure ZB phase InAs$_{1-x}$Sb$_x$ nanowires can be achieved by further increasing the antimony content $x$. Figure 2 shows high resolution transmission electron microscope (HRTEM) images and corresponding selected area electron diffraction (SAED) patterns of <0001>- or <111>-oriented InAs$_{1-x}$Sb$_x$ nanowires. To minimize the effect of nanowire diameter on its crystal structure, the nanowire diameter shown here is fixed in the range from 31 nm to 54 nm. It can be easily found from Figure 2 that the phase purity of InAs$_{1-x}$Sb$_x$ nanowires depends on the antimony content $x$. With the increase of the antimony content $x$, the favorable phase of nanowires gradually changes from WZ to ZB (see Figure 2g-l). When the antimony content $x$ exceeds 0.25 (see Figure 2c-f), all nanowires exhibit a pure ZB phase. These results indicate that the fine control of antimony content is an effective method to grow pure-phase InAs$_{1-x}$Sb$_x$ nanowires.

As shown in Figure 2, there are always a large number of defects in InAs$_{1-x}$Sb$_x$ nanowires with the antimony content $x$ less than 0.2. For many potential applications, such as wide spectrum infrared detectors, it is also important to prepare pure-phase nanowires with low antimony content ($x < 0.2$).[6, 25, 31, 36] As shown in Figure 3, we find that pure-phase InAs$_{1-x}$Sb$_x$ nanowires ($0 \leq x \leq 0.63$) can be achieved by controlling their diameters or growth directions. It can be seen from Figure 3a that the <0001>-oriented InAs$_{0.93}$Sb$_{0.07}$ nanowire with a diameter of 13 nm exhibits pure WZ phase. And the density of planar defects in this nanowire can be as low as 10 $\mu m^{-1}$. It indicates that reducing the diameter can effectively improve the phase purity of InAs$_{1-x}$Sb$_x$ nanowires with a low antimony content. However, this method may not be suitable for nanowires with higher antimony contents. The reasons are as follows: 1) it is difficult to obtain ultrathin nanowires due to the surfactant effect of antimony (see Figure 1b); 2) ZB phase becomes the dominant phase as $x$ increases (see Figure 2). Figure 3b-f shows the HRTEM images and corresponding SAED patterns of InAs$_{1-x}$Sb$_x$ nanowires grown along non-<111> directions. It can be seen that all the InAs$_{1-x}$Sb$_x$ nanowires with different antimony contents grown along <100>, <110> and <112> directions exhibit pure ZB phase (see Figure 3b-f). Similar results have also been found in InAs nanowires.[15, 39-42] It indicates that the control of the growth direction of nanowires is also effective way to obtain pure-phase InAs$_{1-x}$Sb$_x$ nanowires. This method may be easy to implement by using substrates with different crystallographic orientations.[32] Above results confirm that pure-phase InAs$_{1-x}$Sb$_x$ nanowires with a large-composition-range ($0 \leq x \leq 0.63$) can be successfully fabricated by MBE.

**Homogeneous compositional distribution.** In addition to the control of phase purity, the effective control of compositional distribution of ternary nanowires is also very important for tuning their properties. As mentioned above, the unevenly compositional distribution has always been found in InAs$_{1-x}$Sb$_x$ nanowires.[9, 20, 25, 33] However, based on detailed EDS analyses, we find that indium, arsenic and antimony are uniformly distributed in our InAs$_{1-x}$Sb$_x$ nanowires, and no spontaneous core-shell nanostructures are found. Figure 4 shows high-angle annular dark field scanning transmission electron microscope (HAADF-STEM) images and EDS data of InAs$_{1-x}$Sb$_x$ nanowires with different antimony contents. We can see from Figure 4a$_0$-f$_4$ that indium, arsenic and antimony are uniformly distributed for all InAs$_{1-x}$Sb$_x$ nanowires. It is also worth noting that the compositional distribution near the catalyst-nanowire interface exhibits slight inhomogeneity due to the reservoir effect of catalyst (this phenomenon has been discussed in detail in our previous article).[32] In addition, silver-indium catalyst segregation[32, 43] is suppressed with increasing $x$, which may be related to antimony residues in the catalyst nanoparticle. The quantitative EDS data indicate that there are higher antimony residues in the catalyst nanoparticle for those nanowires with a higher antimony content. The high antimony residues can lower the surface energy of the catalyst nanoparticle,[14, 32, 43-44] which makes it stable in the air.

Figure 4g$_0$ is a cross section HAADF-STEM image of the InAs$_{0.71}$Sb$_{0.29}$ nanowire, which shows that the nanowire has the hexagonal cross section (more details available in S2 in the supporting information). We can see from Figure 4g$_1$-4h that the distribution of indium, arsenic and antimony is uniform across the cross section of the nanowire, and no obvious core-shell nanostructures formed. From Figure 4g$_2$, it can be found that the cross section of the EDS maps of antimony is slightly larger than that of arsenic. This phenomenon may be related to the formation of a native oxide layer (see Figure 4g$_3$,h). Quantitative EDS data prove that the variation of the antimony content $x$ on the cross section is less than 2%. Our results are different from those of foreign-catalyst-free or gold-assisted nanowires, where the antimony distribution on the cross section is significantly heterogeneous, and even the spontaneous core-shell structure can be found.[9, 20, 25, 33] In those works, InAs$_{1-x}$Sb$_x$ nanowires show a larger diameter (typically 100 ~ 200 nm) due to the strong vapor-solid growth on the sidewalls of nanowires. In contrast, our nanowires show a smaller diameter (typically 20 ~ 60 nm), and no obvious vapor-solid growth is observed on the sidewalls of nanowire. These results indicate that a small nanowire diameter may be the key to obtain homogeneous InAs$_{1-x}$Sb$_x$ nanowires.

**Electrical properties of InAs$_{1-x}$Sb$_x$ nanowires.** We evaluate the electrical properties of as-grown InAs$_{1-x}$Sb$_x$ nanowires using back-gated field-effect transistor (FET) devices. The schematic diagram and SEM image of the InAs$_{1-x}$Sb$_x$ nanowire device are presented in Figure 5a. In this device, the bias voltage ($V_{ds}$) is applied to the drain electrode, the source electrode is grounded, and the back-gate voltage ($V_g$) is used to tune the carrier concentration of the nanowire. Figure 5b shows the output characteristics of InAs$_{0.93}$Sb$_{0.07}$ nanowire FETs at 300 K with different $V_g$. It can be seen that the device shows good linear shape and current saturation, which suggests the ohmic contact of the device. We use $\sigma = \frac{4GL}{\pi D^2}$ to extract the conductivity $\sigma$ of nanowires, where $G$ is the conductance, $L$ is the distance between the contacts and $D$ is the diameter of nanowire. As shown in Figure 5c, it can be found that InAs$_{1-x}$Sb$_x$ nanowires with higher antimony content show better conductivity for $x \leq 0.29$, which is consistent with the reported results of foreign-catalyst-free nanowires.[21, 25] For $x \geq 0.29$, the average conductivity of nanowires is independent of the antimony content $x$, which may be related to the fact that they have substantially the same bandgap at room temperature.[11]

Figure 5d shows the typical transfer characteristics of InAs$_{1-x}$Sb$_x$ FETs with different antimony contents at $V_{ds}$ = 100 mV. It can be seen that all devices exhibit n-type conduction. The on-off ratios of these devices range from 4 to 22, which indicates InAs$_{1-x}$Sb$_x$ nanowires cannot be turned off completely at 300 K due to their narrow bandgaps.[11, 21, 45] The field-effect mobility of nanowires can be estimated using $\mu = \frac{g_m L^2}{C V_{ds}}$, where $g_m$ is transconductance and $C$ is nanowire-gate capacitance. For back-gated FETs, the nanowire-gate capacitance can be approximated by $C = \frac{2\pi\varepsilon_0\varepsilon_r L}{\cosh^{-1}(1+\frac{2t_{ox}}{D})}$, where $\varepsilon_0$ is permittivity, $\varepsilon_r$ is relative permittivity and $t_{ox}$ is the thickness of the dielectric layer.[46] Using the estimated conductivity and mobility, the carrier concentration $n$ can also be calculated by $n = \frac{\sigma}{e\mu}$, where $e$ is the elementary charge. As shown in Figure 5e, the average mobility of nanowires first increases with the increase of the antimony content, and then tends to be saturated for $x \geq 0.29$. The carrier concentration $n$ varies slightly with the antimony content (see Figure 5f). The maximum mobility ($\mu_m$) of nanowires at 300 K can reach 1400 cm$^2$V$^{-1}$s$^{-1}$, which demonstrates high-quality InAs$_{1-x}$Sb$_x$ nanowires have been grown by MBE using silver catalysts.

Figure 6 shows the output and transfer characteristics of InAs$_{1-x}$Sb$_x$ nanowire FETs at 7 K. We can see that these devices still present good ohmic contacts (see Figure 6a), and most InAs$_{1-x}$Sb$_x$ FETs exhibit n-type conduction (see Figure 6b). Different from the results at 300 K, these devices show higher on-off ratios ($10^3 \sim 10^4$) at 7 K, which suggests the carrier concentration in the junction region can be effectively tuned by $V_g$. In addition, ambipolar behavior is observed in some InAs$_{1-x}$Sb$_x$ nanowires ($x \geq 0.39$). As shown in Figure 6c, as the $V_g$ becomes more negative, the conduction type of the InAs$_{0.44}$Sb$_{0.56}$ nanowire FET changes from n-type to p-type. Ambipolar behavior is usually found in narrow bandgap nanowires, which can be explained by band-to-band tunneling mechanism.[47-51] At large negative $V_g$, the tunneling of hole from the valence band in the junction region to the conduction band in the electrode region results in the formation of a reverse current.

Based on the output and transfer characteristics of nanowire FETs, the mobility and carrier concentration of nanowires at 7 K are also extracted. We can see from Figure 6d that the mobility of nanowires at 7 K increases approximately 3 times compared to the results at room temperature. This is because phonon scattering is significantly suppressed at low temperatures, thereby increasing the mobility of the nanowires. The trend of the mobility varying with the antimony content $x$ at 7 K is similar to that at 300 K, and the maximum mobility of InAs$_{1-x}$Sb$_x$ nanowires with a diameter of 40 ~ 70 nm can be up to 4200 $cm^2V^{-1}s^{-1}$. The estimated mobility of our InAs$_{1-x}$Sb$_x$ nanowires is comparable to that of InAs$_{0.3}$Sb$_{0.7}$/Sn semiconductor-superconductor nanowires (InAs$_{0.3}$Sb$_{0.7}$ nanowires grown on InAs nanowire stems using gold as catalysts), which suggests our nanowires have great potential for achieving high-quality hybrid semiconductor-superconductor nanowire devices.[2] As shown in Figure 6e, unexpectedly, InAs$_{0.93}$Sb$_{0.07}$ nanowires show slightly higher carrier concentration than other InAs$_{1-x}$Sb$_x$ nanowires ($x \geq 0.13$). It may be understood by the diameter-dependent electrical transport properties of nanowires.[40, 52-53] We can see from Figure 6f that InAs$_{1-x}$Sb$_x$ nanowires with the smaller diameter exhibit higher carrier concentration (especially for $D < 30$ nm). Similar phenomena have also been observed in InAs nanowires, where nanowires with a smaller diameter show higher carrier concentration due to the presence of a surface accumulation layer.[40, 52-53] As a result, in our case, it can be inferred that the surface accumulation layer of InAs$_{0.93}$Sb$_{0.07}$ nanowires can contribute more carriers due to their smaller diameters, which makes their carrier concentration higher than that of other samples (see Figure 6f).

In summary, we demonstrate the growth of large-composition-range stemless InAs$_{1-x}$Sb$_x$ nanowires on Si (111) substrates by MBE for the first time. The antimony content $x$ can be tuned from 0 to 0.63, which is about two times larger than the previous experiments ($0 \leq x \leq 0.35$). Furthermore, we also realize the control of phase purity of InAs$_{1-x}$Sb$_x$ nanowires. Pure-phase InAs$_{1-x}$Sb$_x$ nanowires have been achieved by increasing the antimony content, reducing the nanowire diameter and growing nanowires along non-<111> directions. In particular, we resolve the problem of uneven compositional distribution of InAs$_{1-x}$Sb$_x$ nanowires. Detailed EDS analyses confirm that the antimony is uniformly distributed along the axial and radial directions of InAs$_{1-x}$Sb$_x$ nanowires, and no spontaneous core-shell nanostructures are found. In addition, electrical measurements confirm that our InAs$_{1-x}$Sb$_x$ nanowires have high motilities, which further prove that the nanowires have high crystal quality. Our work paves the way toward future high-performance InAs$_{1-x}$Sb$_x$ nanowire electronic, optoelectronic and quantum devices.

**EXPERIMENTAL METHOD**

**Nanowire Growth.** All silver-assisted InAs$_{1-x}$Sb$_x$ nanowires were grown in a solid source MBE system (VG V80H) on Si (111) substrates. Before loading the Si (111) substrate into the MBE chamber, a diluted HF (2%) solution was used to remove its surface contamination and native oxide layer. Prior to nanowire growth, the silver layer with a nominal thickness of less than 0.5 nm was deposited at room temperature and then annealed in situ at 600 °C under constant arsenic flux to generate silver catalyst nanoparticles. The nanowire growth was started by opening indium and antimony shutters simultaneously when the growth temperature reached 505 °C. To obtain InAs$_{1-x}$Sb$_x$ nanowires with different antimony contents, the indium and arsenic fluxes were fixed at $1.6 \times 10^{-7}$ mbar and $1.6 \times 10^{-6}$ mbar, respectively, while the antimony flux was changed from $1.3 \times 10^{-7}$ mbar to $6.4 \times 10^{-6}$ mbar. The nanowire growth lasts 40 minutes. The morphology and crystal structure of InAs$_{1-x}$Sb$_x$ nanowires were investigated by a FEI NanoSEM 650 scanning electron microscope and a JEM F200 transmission electron microscope, respectively. In the JEM F200 system, the compositional distribution of nanowires was also measured by EDS during HAADF-STEM observations.

**Device Fabrication.** For the electrical characterizations of InAs$_{1-x}$Sb$_x$ nanowires, the back-gated FETs were fabricated. The as-grown nanowires were firstly transferred to a highly p-doped Si (100) substrate covered with a 300 nm thermal oxide layer. Then, the metal electrodes were attached by the standard electron beam lithography process. Before depositing metal, a diluted $(NH_4)_2S_x$ solution was used to remove the native oxide layer of nanowires in the contact areas. After that, a chromium/gold (20 nm/80 nm) film was evaporated followed by lift-off in acetone. All electrical measurements of nanowire devices were performed on a probe station using a semiconductor parameter analyzer.

## ACKNOWLEDGMENTS

This work was supported by the National Natural Science Foundation of China (grant nos. 61974138 and 92065106), Beijing Natural Science Foundation (grant no. 1192017) and the Strategic Priority Research Program of Chinese Academy of Sciences (Grant No. XDB28000000). D.P. also acknowledges the support from Youth Innovation Promotion Association, Chinese Academy of Sciences (no. 2017156).

## NOTES

The authors declare that they have no competing financial interests.

**Figures and captions**

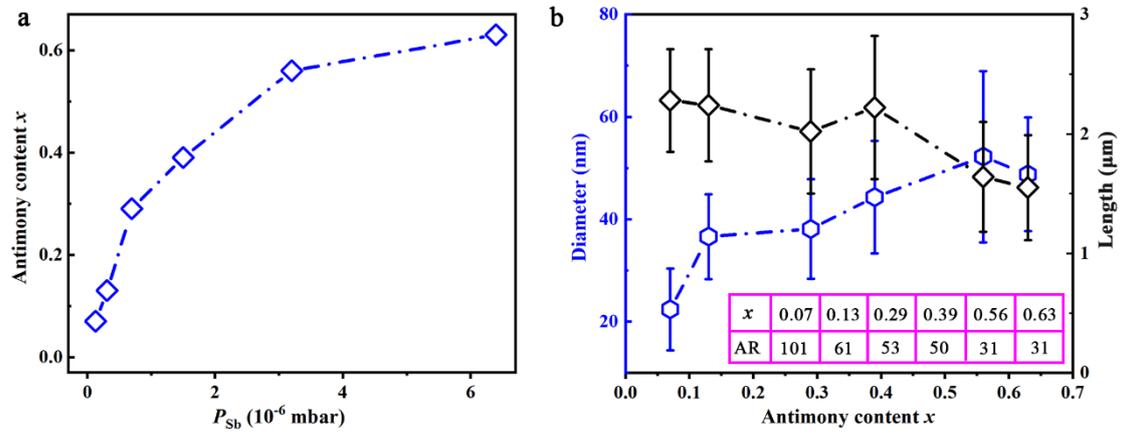

**Figure 1.** a) The antimony content $x$ of InAs$_{1-x}$Sb$_x$ nanowires as a function of the antimony flux $P_{Sb}$. b) The average diameter and length of InAs$_{1-x}$Sb$_x$ nanowires as a function of the antimony content $x$. The AR of nanowires is also given in panel (b). Each data point in panel (b) is a statistical average of more than 50 nanowires.

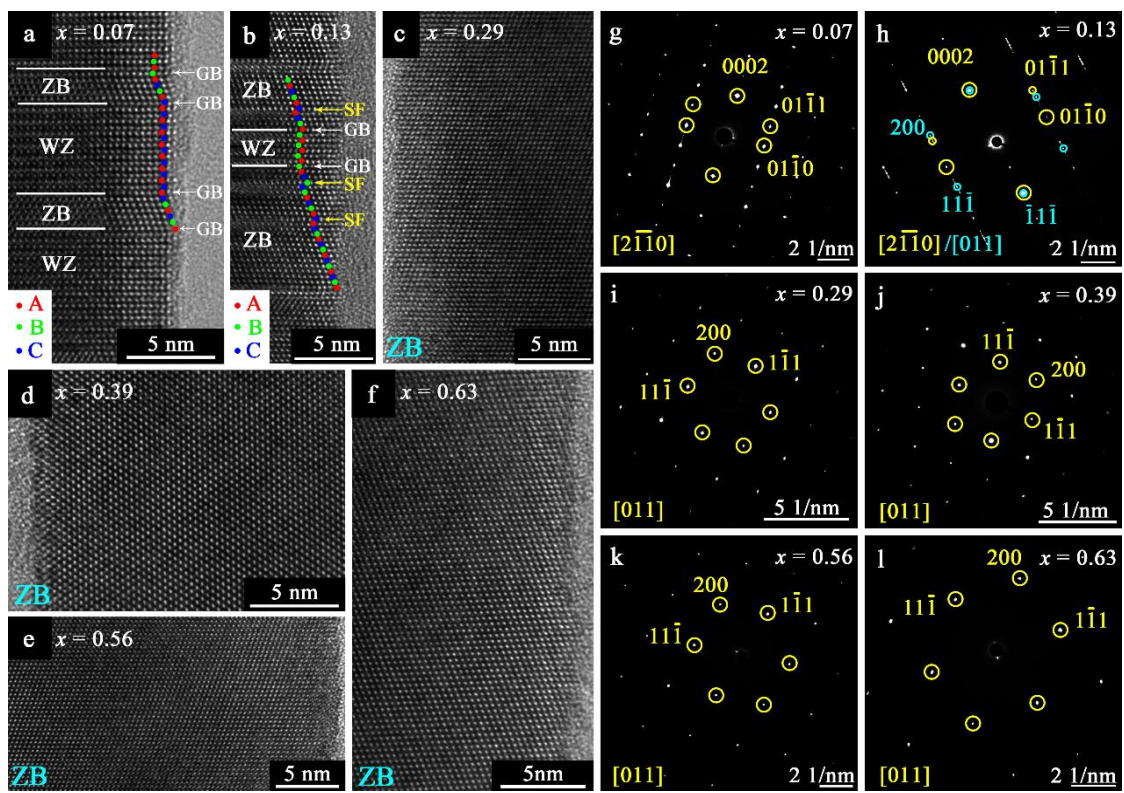

**Figure 2.** a-f) Typical HRTEM images of <0001>- or <111>-oriented InAs$_{1-x}$Sb$_x$ nanowires with $x$ of 0.07, 0.13, 0.29, 0.39, 0.56 and 0.63, respectively. g-l) The corresponding SAED patterns of InAs$_{1-x}$Sb$_x$ nanowires in panel (a-f).

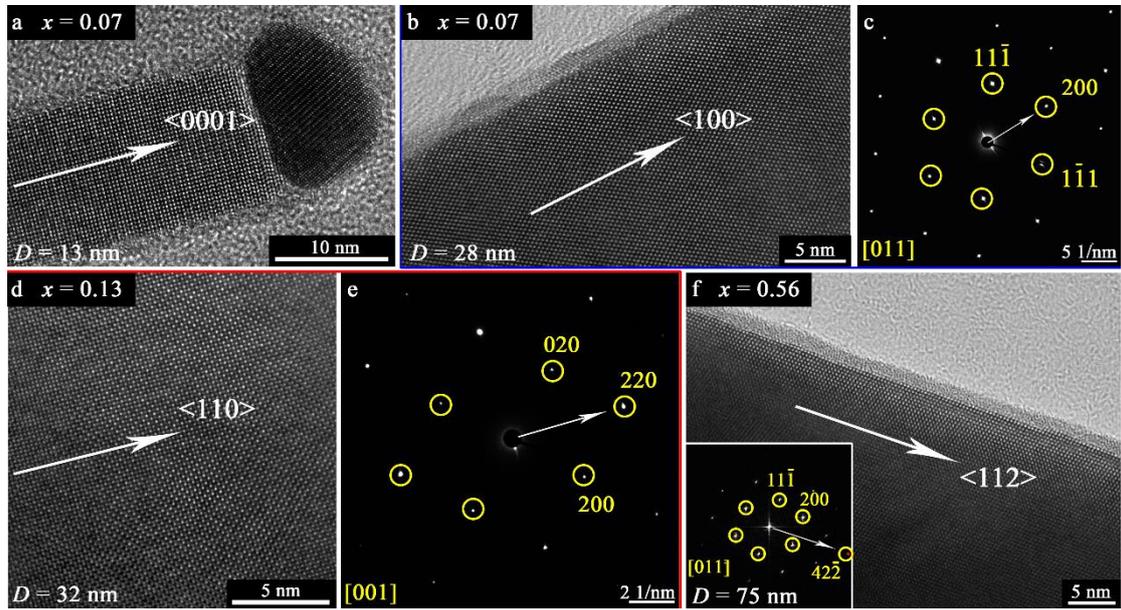

**Figure 3.** a) HRTEM image of the <0001>-oriented InAs$_{0.93}$Sb$_{0.07}$ nanowire with a diameter of 13 nm. b,c) HRTEM image and SAED pattern of the <100>-oriented InAs$_{0.93}$Sb$_{0.07}$ nanowire with a diameter of 28 nm. d,e) HRTEM image and SAED pattern of the <110>-oriented InAs$_{0.87}$Sb$_{0.13}$ nanowire with a diameter of 32 nm. f) HRTEM image of the <112>-oriented InAs$_{0.44}$Sb$_{0.56}$ nanowire with a diameter of 75 nm. The inset is its fast Fourier transform image.

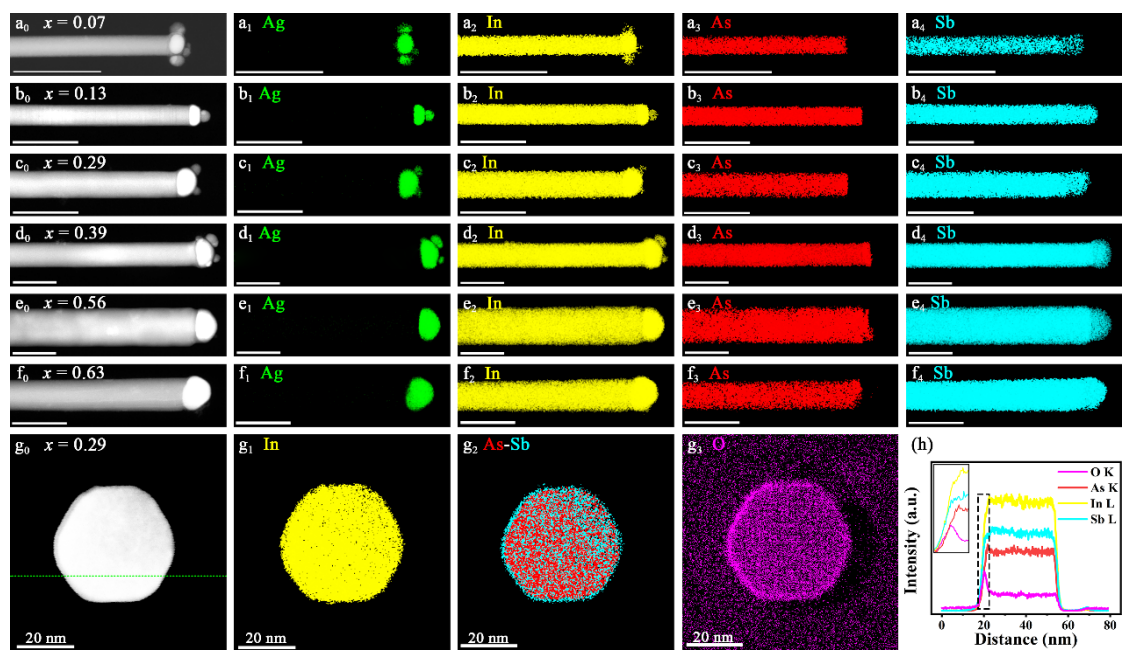

**Figure 4.** $a_0$-$f_4$) HAADF-STEM images and corresponding false-color EDS maps of InAs$_{1-x}$Sb$_x$ nanowires with $x$ of 0.07, 0.13, 0.29, 0.39, 0.56 and 0.63, respectively. $g_0$-$g_3$) The cross section HAADF-STEM image and corresponding false-color EDS maps of the InAs$_{0.71}$Sb$_{0.29}$ nanowire. h) The EDS line scans taken along the green dotted line in panel ($g_0$). Scale bars in panel ($a_0$-$f_4$): 100 nm.

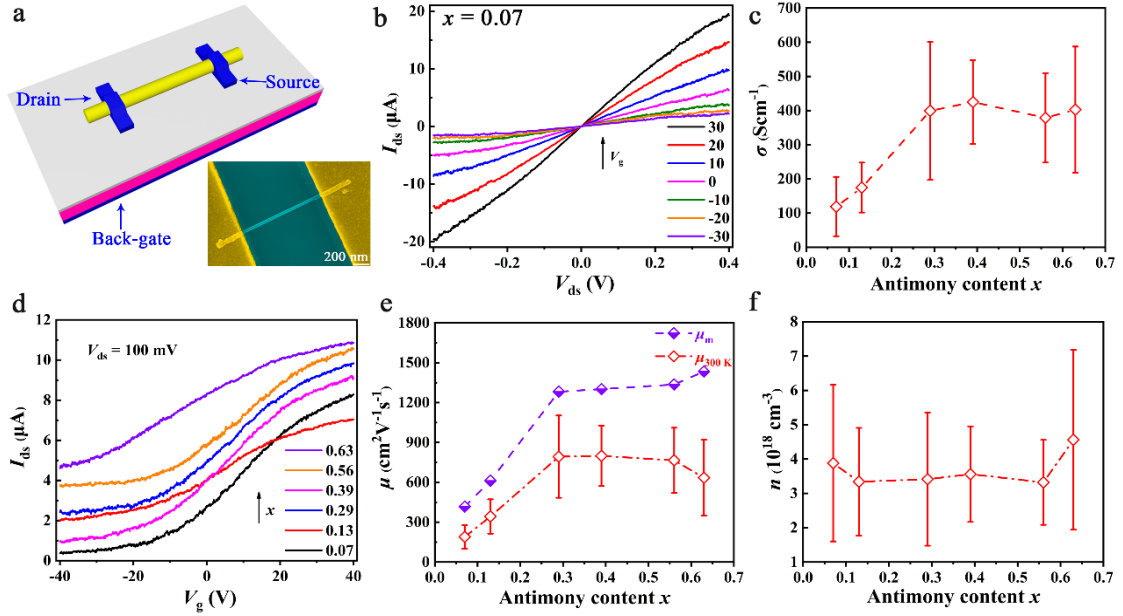

**Figure 5.** a) A schematic diagram and typical SEM image of the InAs$_{1-x}$Sb$_x$ nanowire FET. b) Output characteristics of the InAs$_{0.93}$Sb$_{0.07}$ nanowire FET at 300 K with different $V_g$. c) Average conductivity of nanowires as a function of the antimony content $x$. d) Transfer characteristics of InAs$_{1-x}$Sb$_x$ nanowire FETs at 300 K. e) Average mobility ($\mu_{300\,K}$) and the maximum mobility ($\mu_m$) of nanowires as a function of the antimony content $x$. f) Average carrier concentration of nanowires as a function of the antimony content $x$. Each data point in panel (d,f) is a statistical average of the electrical results of 15-25 nanowires.

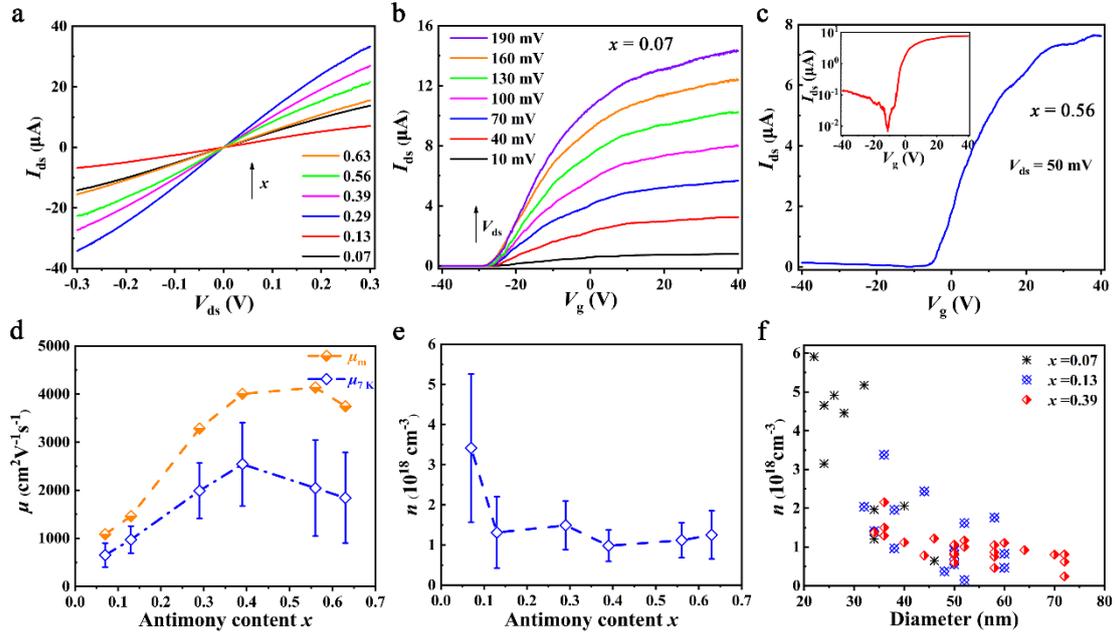

**Figure 6.** a) Output characteristics of InAs$_{1-x}$Sb$_x$ nanowire FETs at 7 K. b,c) Transfer characteristics of the InAs$_{0.93}$Sb$_{0.07}$ and InAs$_{0.44}$Sb$_{0.56}$ nanowire FETs at 7 K, respectively. d) Average mobility ($\mu_{7\,K}$) and maximum mobility ($\mu_m$) of nanowires as a function of the antimony content $x$. e) Average carrier concentration of nanowires as a function of the antimony content $x$. f) Carrier concentration of nanowires with different antimony contents as a function of the diameter. Each data point in panel (d,e) is a statistical average of the electrical results of 10-25 nanowires.

# Supporting Information

# Large-composition-range pure-phase homogeneous InAs$_{1-x}$Sb$_x$ nanowires

Lianjun Wen,[†,‡] Dong Pan,[*,†,‡,⊥] Lei Liu,[†,‡] Shucheng Tong,[†] Ran Zhuo,[†,‡] and Jianhua Zhao[*,†,‡,§]

[†]*State Key Laboratory of Superlattices and Microstructures, Institute of Semiconductors, Chinese Academy of Sciences, P.O. Box 912, Beijing 100083, China*
[‡]*College of Materials Science and Opto-Electronic Technology, University of Chinese Academy of Sciences, Beijing 100049, China*
[⊥]*Beijing Academy of Quantum Information Sciences, 100193 Beijing, China*
[§]*CAS Center for Excellence in Topological Quantum Computation, University of Chinese Academy of Sciences, Beijing 100049, China*

*E-mails: pandong@semi.ac.cn; jhzhao@semi.ac.cn

**Contents**

**S1 Large-composition-range InAs$_{1-x}$Sb$_x$ nanowires grown on Si (111) substrates**

**S2 Compositional distribution in the cross section of InAs$_{1-x}$Sb$_x$ nanowires**

## S1 Large-composition-range InAs$_{1-x}$Sb$_x$ nanowires grown on Si (111) substrates

Figure S1 shows SEM images of InAs$_{1-x}$Sb$_x$ nanowires with different antimony contents grown on Si (111) substrates by MBE. We can see that <111>- and non-<111>-oriented nanowires coexist on the surface of Si (111) substrates. With increasing the antimony content $x$, the density of nanowires begins to gradually decrease. This phenomenon can be understood by the fact that a high antimony flux is detrimental to nanowire growth.[1]

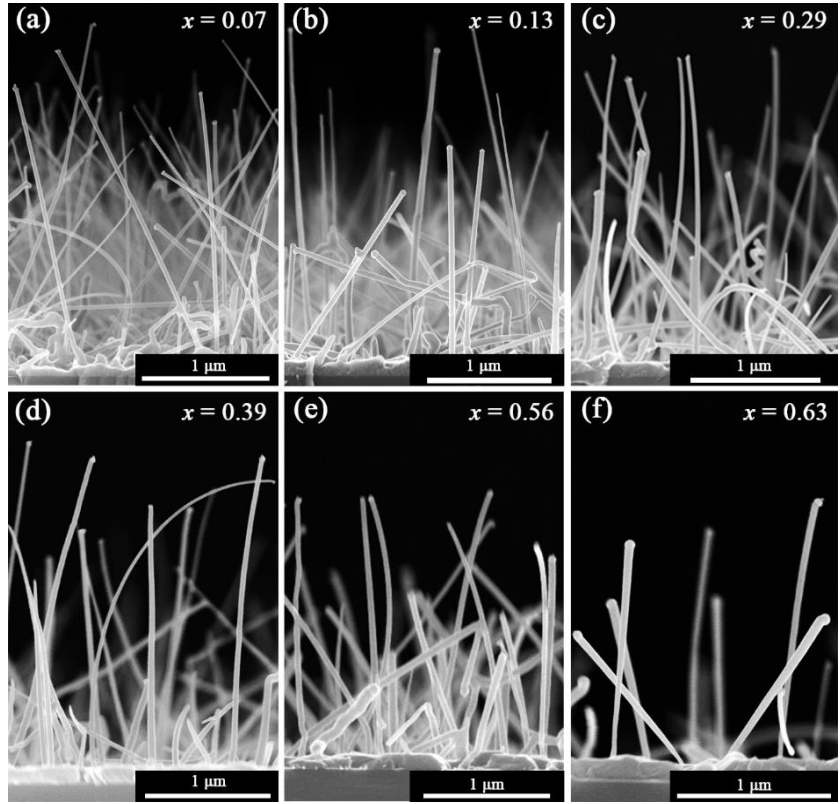

**Figure S1** (a-f) Typical side-view SEM images of InAs$_{1-x}$Sb$_x$ nanowires with the antimony content $x$ of 0.07, 0.13, 0.29, 0.39, 0.56 and 0.63, respectively.

## S2 Compositional distribution in the cross section of InAs$_{1-x}$Sb$_x$ nanowires

Figure S2 shows cross section HRTEM images and corresponding EDS maps of InAs$_{1-x}$Sb$_x$ nanowires. We can see that there are no planar defects in pure ZB phase nanowires. The corresponding EDS maps indicate that the indium, arsenic and antimony are uniform across the cross section, and no obvious vapor-solid growth is observed on the sidewalls of the nanowire.

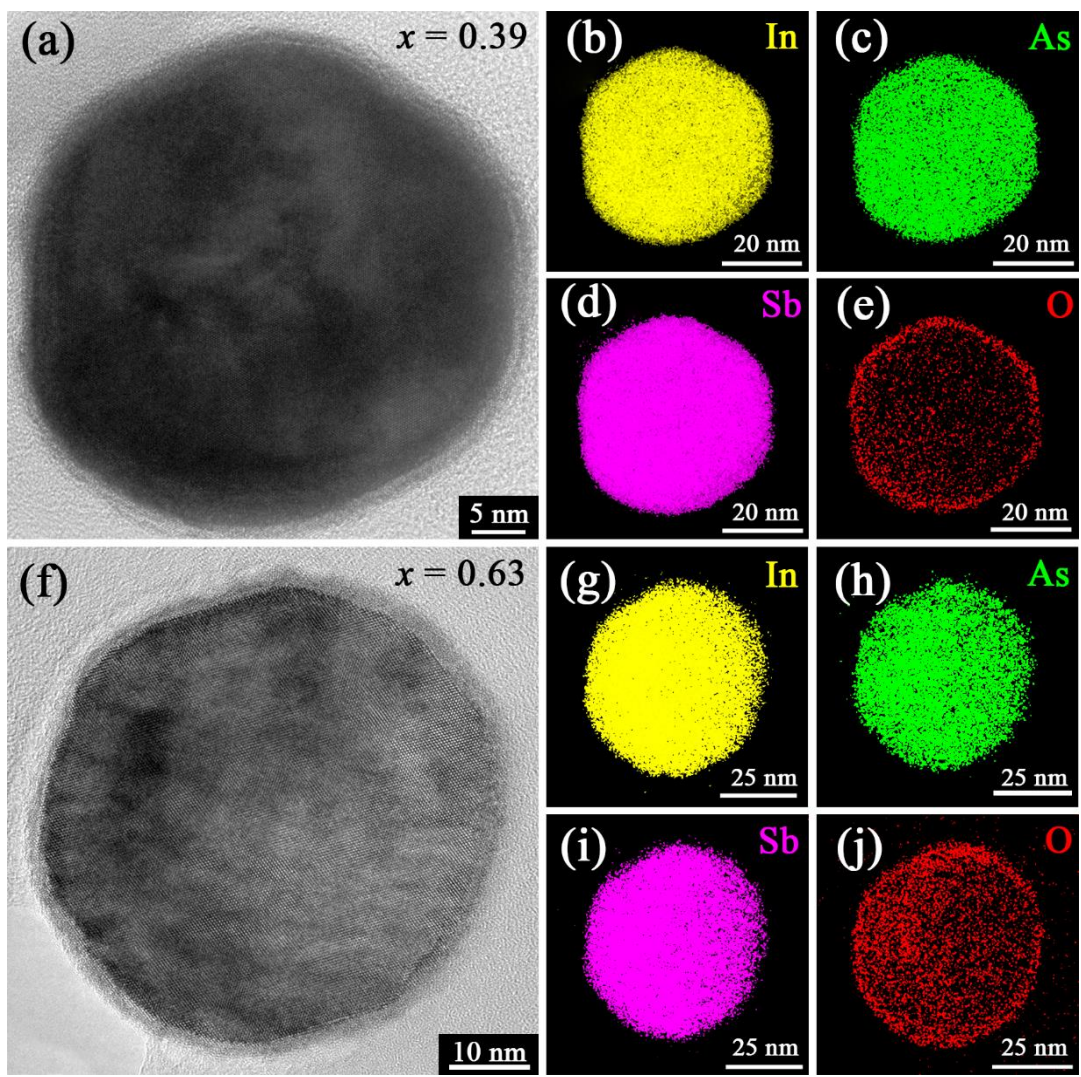

**Figure S2** (a-e) Cross section HRTEM images and EDS maps of the InAs$_{0.61}$Sb$_{0.39}$ nanowire. (f-j) Cross section HRTEM images and EDS maps of the InAs$_{0.37}$Sb$_{0.63}$ nanowire.